\documentclass{webofc}
\usepackage[varg]{txfonts}   

\usepackage{tikz}
\usetikzlibrary{decorations.pathreplacing}

\usepackage{wrapfig}
\usepackage{subcaption}
\usepackage{floatrow}
\newfloatcommand{capbtabbox}{table}[][\FBwidth]

\begin{document}
\title{Quantum Associative Memory in\\ HEP Track Pattern Recognition}

\author{\firstname{Illya} \lastname{Shapoval}\inst{1}\fnsep\thanks{\email{ishapoval@lbl.gov}} \and
        \firstname{Paolo} \lastname{Calafiura}\inst{1}\fnsep\thanks{\email{pcalafiura@lbl.gov}}
}

\institute{Lawrence Berkeley National Laboratory}

\abstract{%
  We have entered the Noisy Intermediate-Scale Quantum Era. A plethora of quantum processor prototypes allow evaluation of potential of the Quantum Computing paradigm in applications to pressing computational problems of the future. Growing data input rates and detector resolution foreseen in High-Energy LHC (2030s) experiments expose the often high time and/or space complexity of classical algorithms. Quantum algorithms can potentially become the lower-complexity alternatives in such cases. In this work we discuss the potential of Quantum Associative Memory (QuAM) in the context of LHC data triggering. We examine the practical limits of storage capacity, as well as store and recall errorless efficiency, from the viewpoints of the state-of-the-art IBM quantum processors and LHC real-time charged track pattern recognition requirements. We present a software prototype implementation of the QuAM protocols and analyze the topological limitations for porting the simplest QuAM instances to the public IBM 5Q and 14Q cloud-based superconducting chips.
}
\maketitle
\section{Introduction} \label{sec:intro}

High Energy Physics (HEP) is a prime example of data intensive science. Over the next decade rapid evolution of accelerator technologies and particle detectors will increase by one order of magnitude the amount and the complexity of data coming from facilities such as Large Hadron Collider (LHC), creating new challenges for the online event selection (Trigger) systems in HEP experiments. To cope with increasing data input rates, sophisticated event selection techniques are being employed at both hardware and software levels. However, current approaches in particular to charged particle pattern recognition, scale poorly with data complexity. Under reasonable technology and cost evolution models, the physics output of the next generation of HEP experiments will be limited by their pattern recognition strategy.

The data input rates foreseen in High-Luminosity LHC (HL-LHC) and beyond impose new challenging requirements on the charged trigger systems. 
Data rates in HEP experiments (at the LHC and elsewhere) will continue to increase. As the corresponding algorithmic complexity of many crucial HEP data processing problems is often polynomial or worse, it is of a substantial interest to investigate alternative, non-classical approaches and algorithms capable of more efficient and scalable track recognition.
To cope with the new challenges, LHC experiments have launched a series of trigger upgrade projects. For example the ATLAS experiment at CERN LHC introduced a new system of electronics, Fast Tracker (FTK) \cite{Shochet_2013}. The system is aimed at real time track reconstruction at a 100 kHz Level-1 trigger rate. To meet the time budget requirements, FTK employed Associative Memory (AM) \cite{Amendolia_1992,Amendolia_1992a}. The latter allowed to address the problem of track pattern recognition -- the most computationally hard part of track finding -- in a massively parallel way. The approach is based on nearly simultaneous and constant-time comparison of coarse-grained hits being readout from the tracker stations to those of the MC generated track patterns pre-loaded into AM. In Run 2 and Run 3, the AM pattern bank will have to store $\sim{}10^{9}$ track patterns of 8-integer
length. The bank pattern of this size requires $8\cdot10^{3}$ AM chips (AMchip06), $\sim{}32 kW$ of supporting power and associated cooling. It is foreseen that 8-16 times more patterns will be required in HL-LHC (2026). There are two evolutionary, linear solutions to this problem, 1) scale-up the number of AM chips of current
generation, considered as cost and power inefficient; 2) more plausibly, upgrade the AMchip06 design to increase its storage capacity.

In this paper, we consider Quantum Associative Memory (QuAM) \cite{Ventura_1998,Ventura_1999,Ventura_2000,Ezhov_2000,Trug_2001} -- a quantum variant of AM based on quantum storage medium and two quantum algorithms for information storage and recall. We compare theoretical QuAM errorless performance expectations to the requirements of the current ATLAS track pattern recognition problem. We also present a software prototype for QuAM circuit generators and point out the limitations for porting QuAM to the state-of-the-art IBM quantum processor units (QPUs).

\section{Quantum storage}\label{sec:storage}

\subsection{Assembling quantum memory}\label{sec:mem_assembling}

Let $\xi \subseteq \{0,1\}^n$ represents a set of $N$ reference binary patterns of length $n$. QuAM is based on establishing an injection $\xi \rightarrowtail \vert \xi \rangle$, where $\vert \xi \rangle$ denotes an orthonormal basis of the Hilbert space of a quantum mechanical system composed of $n$ 2-level qubits. Memorizing $\xi$ can then be done by assembling a quantum superposition:

\[\vert \Xi \rangle = \sum^N_{} \alpha_i\vert \xi^i\rangle ,  \hspace{20pt}  \alpha_i \in \mathbb{C} \hspace{5pt} \wedge \hspace{5pt} N \leqslant 2^n  \hspace{5pt} \wedge \hspace{5pt} \sum^N_{} \vert \alpha_i\vert^2 = 1\]

Note that the special case of $\xi = \{0,1\}^n$ and $N = 2^n$ is trivial and describes complete quantum memory. The only practical value of this case and, more generally, of the case of $N$ approaching $2^n$, is in their setup simplicity, which can be useful for the purposes of verification and benchmarking on the Noisy Intermediate-Scale Quantum Era (NISQ) era QPUs.

Ventura and Martinez~\cite{Ventura_1998, Ventura_1999, Ventura_2000}, and Trugenberger~\cite{Trug_2001}, proposed two alternative ways for unitary assembling of the equal-weighted partial superposition. Simple analysis revealed that the quantum storage algorithm outlined by Trugenberger features shallower circuit and lower topological complexity (see Section~\ref{sec:topo_constrains} for more details), though at the cost of one extra qubit required for auxiliary operations. From the standpoint of limitations of the state-of-the-art QPUs, these properties make this algorithm preferable for implementation.

\begin{figure}[ht!]
\sidecaption
\centering
\includegraphics[width=0.55\columnwidth]{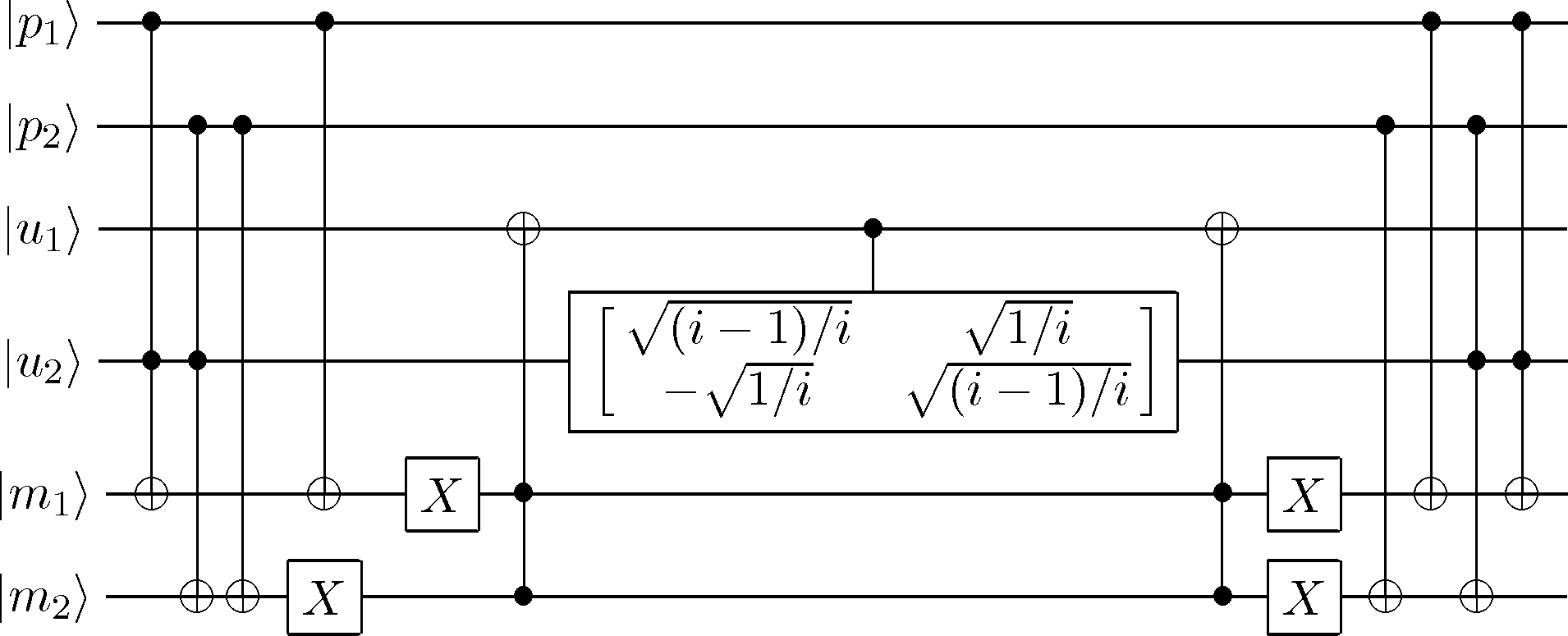}
\caption{The quantum circuit implementing an iteration encoding a 2-bit pattern in the Trugenberger's approach. Qubits $p$ are used as temporary storage register, qubits $u$ - as control register and $m$ - as memory register. The controlled gate, acting on the $u$-register and parameterized with a pattern index $i \in \{1,\dots,N\}$, spawns a new term in the quantum memory superposition for the pattern being stored.} 
\label{fig:iter-store-circuit}
\end{figure}

The Trugenberger's quantum algorithm for storing requires three memory registers spanning $2(n+1)$ qubits to operate~\cite{Trug_2001}: $n$ qubits - for temporary pattern storage, $n$ - for permanent pattern storage and 2 qubits for storage and recall operations control. Figure~\ref{fig:iter-store-circuit} outlines the core iteration for storing a 2-bit pattern in this approach.

A complete quantum circuit for encoding a pattern set must have the iterations interleaved with additional quantum gates for read-in of classical bits into the temporary register. An example of such a complete circuit, along with key implementation details, will be shown in Section~\ref{sec:impl}.

\subsection{Exponential quantum capacity} \label{sec:capacity}

\begin{figure}[ht!]
\vspace{-8pt}
\sidecaption
\centering
\includegraphics[width=0.45\columnwidth]{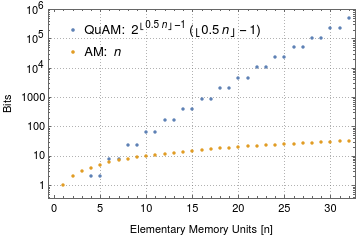}
\caption{Capacity of quantum versus classical associative memories as a function of the number of elementary physical storage units (conventional memory cells in the classical case, and 2-level qubits in the quantum one).}
\vspace{-8pt}
\label{fig:capacity}
\end{figure}

The cardinality of an orthonormal basis of the Hilbert space admits, in computational complexity sense, \emph{optimal} pattern storage capacity for patterns of bit-length equal to the number of qubits in the system. Equivalently, a quantum storage medium provides exponential scaling of its pattern capacity as a function of pattern length. Fair comparison of quantum and classical memory capacities requires accounting for auxiliary qubits that are necessary for quantum operations. However, the asymptotic effect of this additional requirement in storage algorithms of~\cite{Ventura_1998, Ventura_1999, Ventura_2000} and~\cite{Trug_2001} is bound by a constant (in Section~\ref{sec:topo_constrains} we elaborate on other practical consequences of this). Figure~\ref{fig:capacity} compares capacity scaling of classical and quantum associative memories, with the latter being considered in the Trugenberger's approach.

In the case of the ATLAS fast track pattern recognition, the binary pattern length is defined as a sum over the Inner Tracker logical layers of interest $w_{i}$, where $w$ is length of binary representation of a hit identifier within each layer. In LHC Run 2 and Run 3, 8 logical layers are involved in the AM-based pattern recognition. Table \ref{tab:capacity} summarizes available QuAM capacity for various granularity of the track hit resolution.

\begin{table}[h]
\centering
\caption{\label{tab:capacity}QuAM requirements and capacity as a function of width of a detector hit identifier.}

\begin{tabular}[]{@{}lccc@{}}
\hline
Length of detector hit identifier (bits) & 8 & 16 & 32 \\
Length of binary track pattern (bits) & 64 & 128 & 256 \\
QuAM register (qubits) & 130 & 258 & 514 \\
\hline
\textbf{QuAM capacity (patterns)} & \textbf{$\sim$10\textsuperscript{19}} & \textbf{$\sim$10\textsuperscript{38}} & \textbf{$\sim$10\textsuperscript{77}} \\
\hline
\end{tabular}

\end{table}

Qutrits (quantum trits) may help reduce the minimum requirement for the number of physical quantum units necessary for a particular pattern length. But existent hardware implementations are less developed due to more challenging qutrit control. The use of random access encodings~\cite{Ambainis_1999} could further reduce the requirement, though involving a tradeoff on query efficiency. We left both ideas out of the scope of this study. It is important to note, however, that any quantum information compression technique is limited by the Holevo's bound~\cite{Holevo_1973} upon accessing the information.

\section{Quantum recall}\label{sec:recall}

Leveraging the quantum advantage of exponential memory capacity requires scalable and efficient algorithms for memory querying. Two quantum algorithms are discussed in the literature. The first, used by the original proponents of QuAM - Ventura and Martinez~\cite{Ventura_1998, Ventura_1999, Ventura_2000}, is a generalization of the classic Grover's algorithm~\cite{Grover_1996,Grover_1998,Grover_1997}. The algorithm offers quadratic speedup in searching an element in an unordered dataset as compared to the best known classical counterparts, and is proven to be optimal in computational complexity sense~\cite{Zalka_1998}. The second memory querying algorithm \cite{Trug_2001}, relies on the technique of post-selection of the measurement result, and allows to avoid the measurement-induced collapse of memory upon a query. The latter comes at the cost of getting only a binary response. Without a measurement of all pattern bits, a binary response does not provide important features of associative memory such as recall of incomplete and noisy patterns. In addition, the post-selection technique offers no quantum speedup. At the same time, as the algorithm speed is of uttermost importance when working with extremely large memories, we consider the asymptotic speedup - the cornerstone of quantum computing - the guiding feature that makes the Grover's algorithm our primary choice for this study.

\begin{figure}[ht!]
 \centering
 \begin{tikzpicture}
 
 \begin{scope}
 
  \node[inner sep=0pt] (circuit) at (0,-0.5)
    {\includegraphics[trim={6cm 0 0 0},clip, width=0.95\columnwidth]{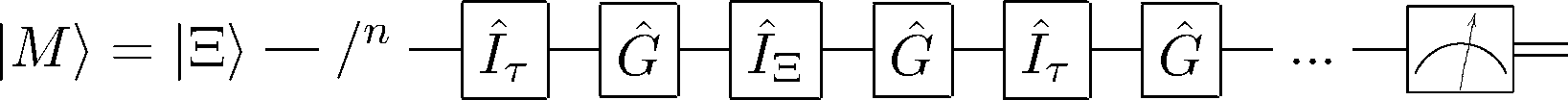}};
    
    \filldraw [rounded corners, color=black!15, fill opacity=0.2] 
       (1.0,-1.2) rectangle (4.5,0.2);
 \end{scope}

\end{tikzpicture}
 \caption{The quantum circuit implementing a variant of the Grover's algorithm generalized for arbitrary (including partial) initial superposition. Only memory register ($n$ qubits) is employed. $\hat{I}_\tau$ is the quantum oracle operator, which inverts the phase of the target state $\tau$. Likewise, $\hat{I}_\Xi$ - inverts phases of all terms originally present in memory. It plays the key role in mitigation the destructive interference of the \emph{ghost} states spawned by the first Grover's diffusion operator. $\hat{G}$ - Grover’s diffusion operator, inverting all amplitudes about their average. $\hat{I}_\tau$ and $\hat{G}$ comprise one Grover iteration. The boxed region denotes the Grover's cycle containing $T_j-2$ (introduced later in formula~\ref{eqn:Grov_iters}) Grover's iterations.}
 \label{fig:retrieve-gen-grover-circuit}
\end{figure}

Application of the Grover's algorithm in the QuAM context requires the use of its generalized variant for handling arbitrary amplitude distributions in the initial memory state \cite{Biron_1999,Biham_1999}. The circuit for  such algorithm is outlined in Figure~\ref{fig:retrieve-gen-grover-circuit}.

In what follows, we revisit some of the theoretical aspects of the algorithm to estimate its {\em theoretical} efficiency in the context of ATLAS FTK pattern recognition requirements.

Let, for a given query, the quantum superposition is split as

\[\vert \Xi \rangle = \sum_{i=1}^{m} k_i(t) \vert \xi^i\rangle +  \sum_{i=m+1}^{N} l_i(t) \vert \xi^i\rangle\]

\noindent with \(k_i(t), l_i(t)\) denoting the amplitudes of matching and non-matching the pattern of interest, \emph{m} -- the number of matched states, and \emph{N} -- the total number of patterns stored in memory. An exact solution to difference equations describing the evolution of amplitudes with arbitrary initial conditions is known \cite{Biron_1999,Biham_1999}. Thus, assuming without loss of generality that \(1\leqslant m \leqslant \dfrac{N}{2}\), the amplitudes of the matching states evolve as

\begin{eqnarray*}
k_i(t) = \sum _{t_j=0}^{t-1} C(t_j) + k_i(0), & \hspace{3pt} C(t) \equiv & \dfrac{2}{N} \left[(N-m)\left(\bar{l}(0)\cos{wt} - \bar{k}(0)\sqrt{\dfrac{m}{N-m}}\sin{wt}\right) \right.\\                     
      & & \left. - m\left(\bar{k}(0)\cos{wt}+\bar{l}(0)\sqrt{\dfrac{N-m}{m}}\sin{wt}\right) \right]
\end{eqnarray*}

\noindent where \(\bar{k}, \bar{l}\) are average amplitudes of matching and non-matching states, and \(\cos{w} = 1 - \dfrac{2m}{N}\). Thus, the probability $\displaystyle P(t) = \sum _{i=1}^m \vert k_i(t) \vert^2$ to measure a marked state (i.e., a solution) peaks at

\begin{equation}
\label{eqn:Grov_iters}
T_j= NI\left(\dfrac{(j+1/2)\pi - \arctan \Bigl(\dfrac{\bar{k}(0)}{\bar{l}(0)} \sqrt{\dfrac{m}{N-m}}\Bigr)} {\arccos(1-\dfrac{2m}{N})}\right),  \quad  j = 0,1,2,\ldots
\end{equation}

\noindent Grover’s iterations, with the nearest integer function {\em NI} defined using the rounding down rule for half-integers.

The upper bound $P_{\max}\ge P(t)$ for probability of measuring a solution is

\begin{equation}
\label{eqn:upper_bound}
P_{\max}=1-\sum_{i=m+1}^N|l_i(t)-\overline{l}(t)|^2\equiv1-(N-m)\sigma_l^2
\end{equation}

\noindent where \(\sigma_l^2\) is variance of non-matching amplitudes, which is a constant of motion \cite{Biron_1999,Biham_1999}. The upper bound (\ref{eqn:upper_bound}) can only be reached for integer arguments of the $NI$ function in (\ref{eqn:Grov_iters}) and equals 1 only in the special case of uniform initial distribution, which can never occur in practical applications of QuAM. However, for $m \ll N$, the theoretical upper bound for measuring a matching state approaches certainty.

\begin{figure}[ht]
\begin{center}
\includegraphics[width=0.4\columnwidth]{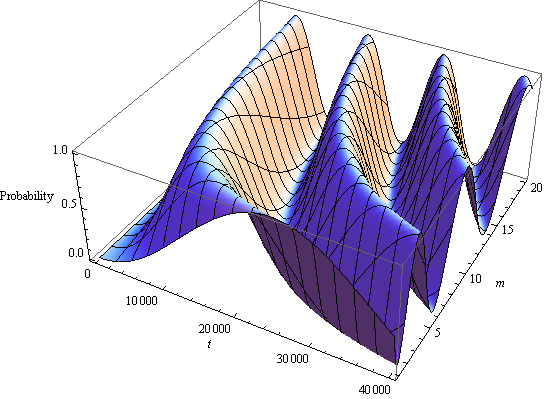}
\includegraphics[width=0.45\columnwidth]{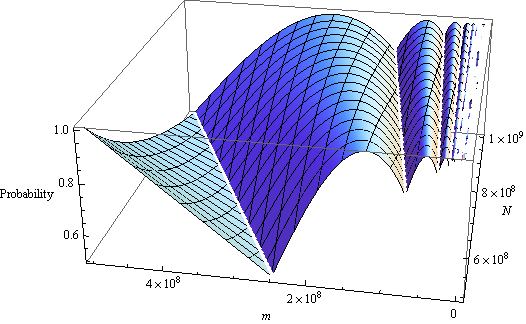}
\end{center}
\caption{Theoretical estimate of (left) probability of measuring a solution as a function of the number of Grover’s iterations and matching patterns (for $N=10^9$) and (right) the peak probability for measuring a solution as a function of the number of matching and total number of patterns stored in QuAM. Both estimates assume the special case of uniform initial superposition and errorless quantum dynamics.}
\vspace{-15pt}
\label{fig:prob}
\end{figure}

As a demonstration, let us consider some of the pertinent properties of the classic Grover's search in the context of the FTK requirements. The evolution of probability as the Grover's algorithm prepares a quantum system for a measurement is visualized in Figure~\ref{fig:prob} (left) for the case of uniform initial superposition of $10^9$ basis states.

\noindent For example, a query matching one pattern requires 24836 Grover's iterations to reach the peak measurement probability of 0.999999999996. Note that the probability ramp-up is slowing down when approaching the peak. This gives an option of cutting the number of iterations down to acceptable value of the outcome probability. Another observation is that, for a given capacity, the number of iterations necessary to reach the peak probability decreases monotonically as the number of solutions increases. For example, a query matching 20 patterns requires 5553 iterations to peak at 0.9999999991 probability of measuring one of the solutions. This suggests, where applicable, another dimension for minimization of the number of Grover iterations based on wildcarded pattern matching. Note that the latter optimization, as seen in Figure~\ref{fig:prob} (right), affects the maximum achievable probability. Significance of this effect, however, is limited to the region of extremely high number of solutions.

\section{Topological constraints} \label{sec:topo_constrains}

Limited connectivity between qubits on most state-of-the-art QPUs often constitutes the main impediment to the mapping of complex quantum algorithms onto them. This can manifest, for example, in non-efficient transpilation of 2-qubit gates leading to higher error accumulation, or in a complete topological mismatch between algorithmic and processor connectivity graphs making the mapping impossible. Connectivity problems can be addressed on the hardware side - with the advancement of the principles of operation and architectures of QPUs toward higher connectivity. They can also be mitigated by optimizing algorithms for lower connectivity requirements. Scalability of such hardware and algorithmic solutions is of uttermost importance, as quantum computing advantages are asymptotic.

\begin{figure}[h!]

\begin{tikzpicture}

\begin{scope}
  \node[inner sep=0pt] (2bit) at (0,0)  {\includegraphics[width=0.16\columnwidth]{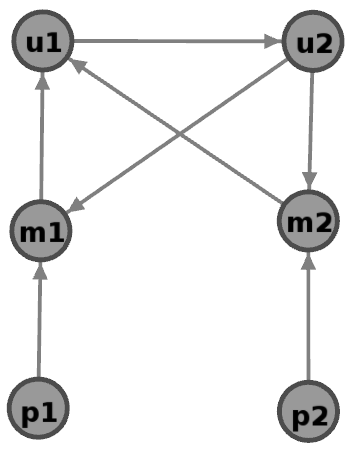}};
  \node[inner sep=0pt] (3bit) at (3.2,0)
  {\includegraphics[width=0.28\columnwidth]{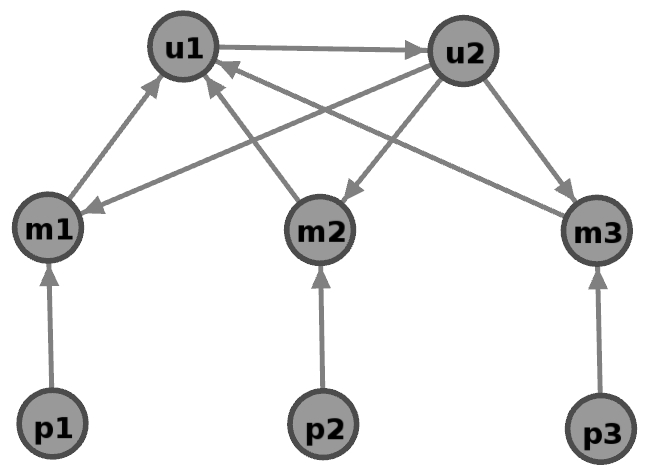}};
\end{scope}

\begin{scope}[xshift=8.5, yshift=0]
  \node[inner sep=0pt] (nbit) at (6.7,-0.07) {\includegraphics[width=0.24\columnwidth]{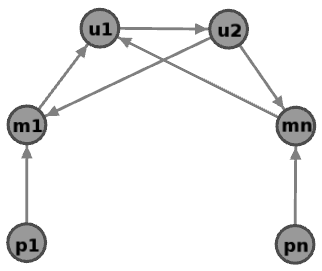}};

  \draw [decorate,decoration={brace,amplitude=6pt,mirror,raise=4ex}]
  (5.2,0.35) -- (8.2,0.35) node[midway,yshift=-3em]{$n$};
  \draw [decorate,decoration={brace,amplitude=6pt,mirror,raise=4ex}]
  (5.2,-0.8) -- (8.2,-0.8) node[midway,yshift=-3em]{$n$};
  
  \path (6.75,0.02) -- node[]{\textbf \ldots} (6.8,0.02);
  \path (6.75,-1.2) -- node[]{\textbf \ldots} (6.8,-1.2);
  
\end{scope}

\end{tikzpicture}
\vspace*{-4mm}
\caption{\label{fig:quam_topology}Integral (storage and recall) topological requirements for (from left to right) 2-bit, 3-bit and $n$-bit patterns. Thus, the connectivity of $p$- and $m$-register qubits is constant (1 and 3, correspondingly) as a function of pattern length, while is linear ($n+1$) for $u$-register qubits.}
   
\end{figure}
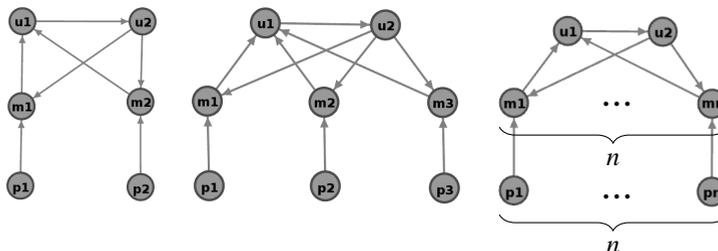

In this light, it is interesting to analyze the topological complexity of the algorithms used in QuAM. It turns out that the storage algorithm suggested by Trugenberger~\cite{Trug_2001} features weaker topological requirements as compared to the original one proposed by Ventura nad Martinez~\cite{Ventura_1998}. The topological requirements for Trugenberger's storage are a superset of the ones for Grover's recall. Thus, we can summarize the integral topological requirements in Figure~\ref{fig:quam_topology}, where we outlined the special cases of 2- and 3-bit patterns, as well as the general case of n-bit patterns. Importantly, the topological complexity of the algorithms does not depend on the number of patterns being stored, but rather on the length of a pattern.

The public IBM Q Experience QPUs we have looked at in this study include, at the time the paper is written, the 5-qubit IBM Q 5 Yorktown/Tenerife and the 14-qubit IBM Q 14 Melbourne~\cite{IBMQ_devices} devices. By the number of qubits, only the latter could allow to run the simplest QuAM circuits (patterns of up to 5-bit length). Unfortunately, it does not satisfy the topological requirements even for the most trivial case of a 2-bit pattern. In contrast, the latter case should be topologically compliant with the 20-qubit IBM Q 20 Tokyo available for IBM Q Network clients and will be tested in the near future.

\section{Implementation on IBM Q Experience} \label{sec:impl}

We have implemented a software prototype that includes the QuAM storage and recall circuit generators. The prototype is based on the Trugenberger’s algorithm for storage and the generalized Grover's algorithm for recall outlined in previous sections and is developed in the QISKit framework~\cite{QISK} of the IBM Q Experience Project~\cite{IBMQ}.

It turns out that most gates employed by both algorithms are either directly implemented, or, like the Toffoli gate, have known exact decompositions over the elementary gate set implemented on IBM Qs. The only minor exception to this is the gate that spawns a superposing term for each new pattern (see the core gate of the storage circuit in Figure~\ref{fig:iter-store-circuit}). The corollary of the \emph{Z-Y} theorem for decomposition of a controlled-\emph{U} operation as $U=e^{i\alpha}(AB)^{-1}\sigma_xB\sigma_xA$~\cite{Nielsen_2011_QCQ} with a choice of $A=U_3(\pi,\theta,0)$ and $B=H$, as well as a substitution of the pattern index $i=\sin^{-2}\frac{\theta}{2}$, make its decomposition straightforward.

\begin{figure}[ht!]
\centering

\vspace{-10pt}

\begin{subfigure}[b]{\columnwidth}
\includegraphics[trim={0 2.3cm 11.5cm 2.6},clip,width=1.0\columnwidth]{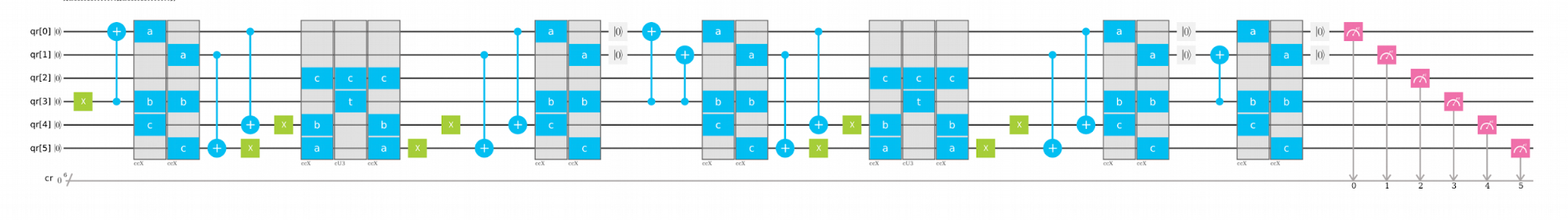}
\end{subfigure}

\begin{subfigure}[b]{\columnwidth}
\includegraphics[trim={2.0cm 0 0 0},clip,width=1.0\columnwidth]{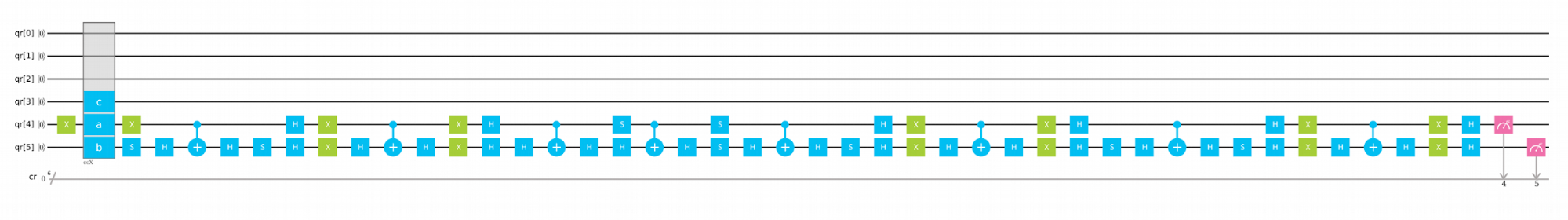}
\end{subfigure}

\vspace{-10pt}

\caption{The end-to-end QuAM circuits prodeced by the Qiskit-based circuit generator prototype for (top): saving three 2-bit patterns and (bottom): retrieving one 2-bit pattern.}
\label{fig:qiskit-circuit}
\end{figure}

Figure~\ref{fig:qiskit-circuit} shows an example of the end-to-end 2-bit quantum circuits produced by our prototype.

\section{Conclusion}
Leveraging the power of the quantum computing paradigm in HEP, and elsewhere, is in its infancy. The objective of this study was to initiate a discussion within the HEP community about the feasibility of applications of QuAM for charged track pattern recognition to the next generation HEP experiments.

In this work, we analyzed the topological limitations of the two QuAM initialization variants and pointed out that, with limited QPU connectivity, implementation of the Trugenberger's algorithm is more feasible on the state-of-the-art IBM QPUs. We evaluated some of the pertinent properties of the generalized Grover's search, extended by Ventura and Martinez, in the context of current and future HEP data processing requirements. We have also prototyped the Trugenberger's initialization and the Grover's algorithm generalized for arbitrary (including partial) initial superpositions in the Qiskit framework, yielding recall probabilities that matched the theoretical estimates up to the machine epsilons. The prototype will allow us to run the simplest instances of QuAM on the IBM Q 20 chip, as well as to simulate the instance of 15-bit pattern QuAM on IBM Q 32 QASM Simulator.

Many important questions, that can ultimately affect the viability of QuAM, are beyond the scope of this paper and will be addressed in follow-up studies. Our next steps will be to scale-up the simulations of QuAM to higher-order patterns and to evaluate its timing and efficiency on state-of-the-art QPUs. We will try to address some QuAM limitations, including linear memory initialization, efficiency of probabilistic state cloning in the Grover-based setting, as well as implement the alternative querying mechanism based on the post-selection of the measurement result.

\section*{Acknowledgement}
This work was funded by DOE HEP Center for Computational Excellency. Our prototype was tested on the IBM Q Experience cloud service.

\end{document}